\documentstyle[12pt]{article}
\begin{document}
\title{{\bf Novel cloning machine with supplementary information }}
\author{Daowen Qiu\\
\small{Department of Computer Science, Zhongshan University,
Guangzhou 510275,}\\{\small People's Republic of China}\\
{\small E-mail address: issqdw@mail.sysu.edu.cn}}
\date{  }

\maketitle
\begin{center}
\begin{minipage}{130mm}
\begin{center}{\bf Abstract}\end{center}
{\small Probabilistic cloning was first proposed by  Duan and Guo.
Then Pati established a {\it novel cloning machine} (NCM) for
copying superposition of multiple clones simultaneously. In this
paper, we deal with the {\it novel cloning machine with
supplementary information} (NCMSI). For the case of cloning two
states, we demonstrate that the optimal efficiency of the NCMSI in
which the original party and the supplementary party can perform
quantum communication equals that achieved by a two-step cloning
protocol wherein classical communication is only allowed between
the original and the supplementary parties. From this equivalence
it follows that NCMSI may increase the success probabilities for
copying. Also, an upper bound on the unambiguous discrimination of
two nonorthogonal pure product states is derived. Our
investigation generalizes and completes the results in the
literature.}


\end{minipage}
\end{center}
\vskip 10mm

\section*{1. Introduction }

\hskip 6mm Over the past decade, quantum computation and quantum
information has been given extensively attention due to the more
power in essence than classical computation [1]. While the
characteristics of quantum principles such as quantum
superposition and entanglement  essentially enhance the power of
quantum information processing, the unitarity and linearity of
quantum physics also lead to some impossibilities---the {\it
no-cloning theorem} [2,3,4] and the {\it no-deleting principle}
[5]. The linearity of quantum theory makes an unknown quantum
state unable to be perfectly copied [2,3] and deleted [5], and two
nonorthogonal states are not allowed to be precisely cloned and
deleted as a result of the unitarity [4,6,7], that is, for
nonorthogonal pure states $|\psi_{1}\rangle$ and
$|\psi_{2}\rangle$, no physical operation in quantum mechanics can
exactly achieve the transformation $|\psi_{i}\rangle\rightarrow
|\psi_{i}\rangle|\psi_{i}\rangle$ $(i=1,2)$. This has been
generalized to mixed states and entangled states [8,9].
Remarkably, these restrictions provide a valuable resource in
quantum cryptography [10], because they forbid an eavesdropper to
gain information on the distributed secret key without producing
errors.

Recently Jozsa [11] and Horodecki {\it et al.} [12] further
clarified the no-cloning theorem and the no-deleting principle
from the viewpoint of conservation of quantum information, and in
light of this point of view two copies of any quantum state
contain more information than one copy; in contrast, two classical
states have only the same information as any one of the two
states. Specifically, Jozsa [11] verified that if supplementary
information, say a mixed state $\rho_{i}$ is supplemented, then
there is a physical operation
\begin{equation}
|\psi_{i}\rangle\otimes\rho_{i}\rightarrow
|\psi_{i}\rangle|\psi_{i}\rangle
\end{equation}
if and only if there exists physical operation
\begin{equation}\rho_{i}\rightarrow |\psi_{i}\rangle,\end{equation}
where by physical operation we mean a completely positive
trace-preserving map, and $\{|\psi_{i}\rangle\}$ is any given
finite set of pure states containing no orthogonal pairs of
states. This result implies that the supplementary information
must be provided as the copy $|\psi_{i}\rangle$ itself, since the
second copy can always be generated from the supplementary
information, independently of the original copy. Therefore, this
result may show the ``permanence" of quantum information; that is,
to get a copy of quantum state, the state must already exist
somewhere. Notwithstanding, cloning quantum states with a limited
degree of success has been proved always possibly. A natural issue
is that if the supplementary information is added in a {\it novel
cloning machine} (NCM) by Pati [13], then whether the optimal
efficiency of the machine may be increased. This problem will be
positively addressed in this paper.

Let us briefly recall the pioneers' works regarding quantum
cloning, and the more detailed references may be referred to
Fiur$\acute{a}\check{s}$ek [14] therein. In general, there are two
kinds of cloners. One is the {\it universal quantum-copying
machine} (UQCM) firstly introduced by Bu\u{z}ek and Hillery [15],
and this kind of machines is deterministic and does not need any
information about the states to be cloned, so it is {\it
state-independent}. To be more precise, the UQCM obtained by
Bu\u{z}ek and Hillery [15] is described by the following unitary
transformation $U$:
\begin{eqnarray}
&& |0\rangle_{a}|Q\rangle_{x}\rightarrow
\sqrt{\frac{2}{3}}|00\rangle_{ab}|\uparrow\rangle+\sqrt{\frac{1}{3}}|+\rangle_{ab}|\downarrow\rangle,\\
&&|1\rangle_{a}|Q\rangle_{x}\rightarrow
\sqrt{\frac{2}{3}}|11\rangle_{ab}|\downarrow\rangle+\sqrt{\frac{1}{3}}|+\rangle_{ab}|\uparrow\rangle,
\end{eqnarray}
where $|Q\rangle_{x}$ is the state of the copying device
(auxiliary state),  $|\uparrow\rangle$ and $|\downarrow\rangle$
are an orthonormal basis states, and
$|+\rangle_{ab}=\frac{1}{\sqrt{2}}(|10\rangle_{ab}+|01\rangle_{ab})$.
The ``universal" means that for any pure state
$|s\rangle_{a}=\alpha |0\rangle_{a}+\beta|1\rangle_{a}$ to be
cloned, the distances
$D_{a}=Tr[\rho_{a}^{(out)}-\rho_{a}^{(id)}]^{2}$, and,
$D_{ab}=Tr[\rho_{ab}^{(out)}-\rho_{ab}^{(id)}]^{2}$ are
independent of $\alpha$, that is to say, the efficiency of cloning
under these measures does not rely on the original state
$|s\rangle_{a}$, where by denoting
$|\Psi\rangle_{abx}^{(out)}=U(|s\rangle_{a}|Q\rangle_{x})$,  then
density operator $\rho_{abx}^{(out)}=|\Psi\rangle_{abx\hskip 5mm
abx}^{(out)\hskip 1mm (out)}\langle \Psi|$, the real output in the
 system $ab$ is  $\rho_{ab}^{(out)}=Tr_{x}[\rho_{abx}^{(out)}]$,
the real output in system $a$ is
$\rho_{a}^{(out)}=Tr_{b}[\rho_{ab}^{(out)}]$; by contrast, the
ideal output in the system $ab$ is
$\rho_{ab}^{(id)}=\rho_{a}^{(id)}\bigotimes\rho_{b}^{(id)}$, where
$\rho_{a}^{(id)}=|s\rangle_{a\hskip 2mm a}\langle s|$,
$\rho_{b}^{(id)}=|s\rangle_{b\hskip 2mm b}\langle s|$, in which
$|s\rangle_{b}=\alpha |0\rangle_{b}+\beta|1\rangle_{b}$. (A direct
calculation shows that $D_{a}=\frac{1}{18}$ for the above UQCM.)
To date many authors have deeply dealt with this kind of cloning
devices (for example, [16-26]). By the way, recently the universal
quantum deleting machines have also been considered [27,28].

The other kind of cloners is {\it state-dependent}, since it needs
some information from the states to be cloned. Furthermore, this
kind of cloning machines may be divided into three fashions of
cloning: First is probabilistic cloning machines proposed firstly
by Duan and Guo [29,30], and then by Chefles and Barnett [31] and
Pati [13], and Han {\it et al.} [32], that can clone linearly
independent states with nonzero probabilities. Duan and Guo's
machine can be stated as follows: For states secretly chosen from
the set
$S=\{|\psi_{1}\rangle,|\psi_{2}\rangle,\ldots,|\psi_{n}\rangle\}$,
there is unitary operator $U$ such that
\begin{equation}
U(|\psi_{i}\rangle|\Sigma\rangle|P_{0}\rangle)=\sqrt{r_{i}}|\psi_{i}\rangle|\psi_{i}\rangle|P_{0}\rangle+
\sum_{j=1}^{n}c_{ij}|\Phi_{AB}^{(j)}\rangle|P_{j}\rangle, \hskip
10mm (i=1,2,\ldots,n),
\end{equation}
if and only if states
$|\psi_{1}\rangle,|\psi_{2}\rangle,\ldots,|\psi_{n}\rangle$ are
linearly independent, where $r_{i}$ is the probability of success
for copying $|\psi_{i}\rangle$, $|\Sigma\rangle$ is a blank state,
$|P_{0}\rangle, |P_{1}\rangle,\ldots,|P_{n}\rangle $ are probe
states and orthonormal, and $|\Phi_{AB}^{(j)}\rangle$ are $n$
normalized states of the composite system $AB$.  Therefore, a
general unitary evolution together with a post-selection by
measurement results, yields faithful copies of the input states
with certain probabilities. Indeed, a more general unitary
evolution of the system $ABP$ can be decomposed as the form:
\begin{equation}
U(|\psi_{i}\rangle|\Sigma\rangle|P_{0}\rangle)=\sqrt{r_{i}}|\psi_{i}\rangle|\psi_{i}\rangle|P^{(i)}\rangle
+\sqrt{1-r_{i}}|\Phi_{ABP}^{(i)}\rangle,\hskip 10mm
(i=1,2,\ldots,n),
\end{equation}
that can be stated as: The states
$|\psi_{1}\rangle$,$|\psi_{2}\rangle$,$\ldots$,$|\psi_{n}\rangle$
can be probabilistically cloned with efficiencies $r_{i}$ if and
only if the matrix
$X^{(1)}-\sqrt{\Gamma}X_{P}^{(2)}\sqrt{\Gamma^{+}}$ is positive
semidefinite, where matrices $X^{(1)}=[\langle
\psi_{i}|\psi_{j}\rangle]$, $\sqrt{\Gamma}={\rm diag}
(r_{1},r_{2},\ldots,r_{n})$, $X_{P}^{(2)}=[\langle
\psi_{i}|\psi_{j}\rangle^{2}\langle P^{(i)}|P^{(j)}\rangle]$;
$|P_{0}\rangle$, $|P^{(i)}\rangle$ are normalized states of the
probe $P$ (not generally orthogonal) and
$|\Phi_{ABP}^{(i)}\rangle$ are $n$ normalized states of the
composite system $ABP$ (not generally orthogonal, but it is
required that $\langle P^{(i)}|\Phi_{ABP}^{(j)}\rangle=0$ for any
$i,j=1,2,\ldots,n$). The success probabilities $r_{i}$ and $r_{j}$
satisfy that
\begin{equation}
\frac{r_{i}+r_{j}}{2}\leq \frac{1}{1+|\langle
\psi_{i}|\psi_{j}\rangle|},
\end{equation}
where $|\langle \psi_{i}|\psi_{j}\rangle|\not=1$ is assumed.

Second is deterministic cloners first investigated by Bru\ss\hskip
1mm {\it et al.} [33] and then by Chefles and Barnett [34]. Such a
deterministic cloning machine is described by the unitary operator
$U$:
\begin{equation}
U(|\psi_{i}\rangle^{\otimes M} |\Sigma\rangle^{\otimes
(N-M)})=|\alpha_{i}\rangle,\hskip 10mm (i=1,2,\ldots,n),
\end{equation}
where $|\Sigma\rangle$ is a blank state and $|\alpha_{i}\rangle$
are the output states cloned. According to [33] the global
fidelity $F$ of this cloning device can be expressed as:
\begin{equation}
F=\sum_{i=1}^{n}p_{i}|\langle\alpha_{i}|\psi_{i}\rangle^{\otimes
N}|^{2},
\end{equation}
where $p_{i}$ is the priori probability of the state
$|\psi_{i}\rangle^{\otimes M}$ chosen. From [33,34] it follows
that the optimal output state $|\alpha_{i}\rangle$ must lie in the
subspace spanned by the exact clones $|\psi_{1}\rangle^{\otimes
N},|\psi_{2}\rangle^{\otimes N},\\
\ldots,|\psi_{n}\rangle^{\otimes N}$.

Third is hybrid cloner studied by Chefles and Barnett [32], that
combines deterministic cloner with probabilistic one. The basic
process of cloning is that firstly the initial states, say
$|\psi^{1}_{1}\rangle$ and $|\psi^{1}_{2}\rangle$, are separated
with certain probability $P_{S}$, i.e., a non-unitary
transformation makes with certain probability $P_{S}$ the states
$|\psi^{1}_{1}\rangle$ and $|\psi^{1}_{2}\rangle$ become states
$|\phi_{1}\rangle$ and $|\phi_{2}\rangle$ [31], such that
\begin{equation}
|\langle \phi_{1}|\phi_{2}\rangle|\leq |\langle
\psi_{1}^{1}|\psi_{2}^{1}\rangle|.
\end{equation}
Such a transformation is implemented by some linear operators
$A_{Sk}$ and $A_{Fk}$ satisfying
\begin{equation}
\sum_{k}(A_{Sk}^{\dagger}A_{Sk}+A_{Fk}^{\dagger}A_{Fk})=\hat{{\bf
1}},
\end{equation}
where $\hat{{\bf 1}}$ is identity operator, and
\begin{eqnarray*}
&&A_{Sk}|\psi_{i}^{1}\rangle=s_{ki}|\phi_{i}\rangle,\\
&&A_{Fk}|\psi_{i}^{1}\rangle=f_{ki}|\phi_{i}\rangle,
\end{eqnarray*}
for $i=1,2$, where
\begin{equation}
P_{S}=\sum_{i=1}^{2}\frac{1}{2}\sum_{k}|s_{ki}|^{2}\leq
\frac{1-|\langle \psi_{1}^{1}|\psi_{2}^{1}\rangle|}{1-|\langle
\phi_{1}|\phi_{2}\rangle|}.
\end{equation}
Whereafter, by utilizing deterministic cloner for copying the
states $|\phi_{1}\rangle$ and $|\phi_{2}\rangle$, the states
$|\psi^{2}_{1}\rangle$ and $|\psi^{2}_{2}\rangle$ are
determinately obtained. Therefore, such a cloning scheme obtain
the appropriate states $|\psi^{2}_{i}\rangle$ for copying
$|\psi^{1}_{i}\rangle$ $(i=1,2)$. (Notably, these quantum cloning
machines stated above have been applied to many quantum
cryptographic protocols [35-37].)

The probabilistic machine by Duan and Guo [29,30] can be thought
of as $|\psi\rangle\rightarrow |\psi\rangle^{\otimes 2}$ cloning.
A question addressed by many authors is that given a quantum
state, whether it is possible for a device to produce
$|\psi\rangle\rightarrow |\psi\rangle^{\otimes 2}$,
$|\psi\rangle\rightarrow |\psi\rangle^{\otimes 3}$, $\ldots$,
$|\psi\rangle\rightarrow |\psi\rangle^{\otimes (m+1)}$, in a
deterministic or probabilistic way. Motivated by this proposal and
the idea of probabilistic cloning, Pati [13] established a NCM
that could produce $|\psi\rangle\rightarrow |\psi\rangle^{\otimes
(m+1)}$ $(m=1,2,\ldots,k)$ clones simultaneously, which appear in
a linear superposition of all possible multiple copies with
respective probabilities. Therefore, Pati's NCM [13] generalizes
Duan and Guo's cloning machine [29,30]. For avoiding repetition,
we will describe the NCM in Sections 2 and 3 in detail, and
differentiate between our results and the previous those related.
In this paper, we deal with the {\it NCM with supplementary
information} (NCMSI), and present an equivalent characterization
of such a quantum cloning device in terms of a two-step cloning
protocol in which the original and the supplementary parties are
only allowed to communicate with classical channel.

The remainder of the paper is organized as follows. In Section 2,
we first introduce the existing results regarding probabilistic
cloning with supplementary information, and then present our main
contributions concerning NCMSI. Section 3 is the detailed
demonstration of our major outcomes. In this section, we first
provide a number of related unitary transformations describing
cloning machines, and the corresponding inequalities
characterizing the existence of these unitary transformations are
then given; afterwards, we prove the main results expressed by
Theorem 1 and Theorem 2. Also we derive an upper bound for
unambiguous discrimination of the set
$\{|\psi_{1}\rangle|\phi_{1}\rangle,|\psi_{2}\rangle|\phi_{2}\rangle\}$
(Remark 1). Finally, in Section 4 we summarize our results
obtained, mention some potential of applications, and address a
number related issues for further consideration.

In addition, though some transformations describing cloning
machines have been introduced in Section 1, in the interest of
readability, we would like to present partially them again with
somewhat different forms in Sections 2 and 3 to lead to our
results.

\section*{2. Preliminaries and main results}

\hskip 6mm In this section, we first give the existing results by
Azuma {\it et al.} [38], and then present our main results.

As pointed out above, Jozsa [11] and Horodecki {\it et al.} [12]
verified the no-cloning theorem and the no-deleting principle by
utilizing supplementary information and conversation of quantum
information, respectively. Then we may naturally address that if
supplementary information is added in the NCM, then whether the
success probability for copying will be increased. Recently, Azuma
{\it et al.} [38] suggested probabilistic cloning with
supplementary information by combining probabilistic cloning and
supplementary information. Specifically, for any two
non-orthogonal states $|\psi_{1}\rangle$ and $|\psi_{2}\rangle$,
and supplementary states $|\phi_{1}\rangle$ and
$|\phi_{2}\rangle$, Azuma {\it et al.} [38] showed the following
implication: If there exists unitary operator $U$:
\begin{equation}
U(|\psi_{i}\rangle|\phi_{i}\rangle|P_{0}\rangle)=\sqrt{r_{i}}|\psi_{i}\rangle^{\otimes
(m+1)}|P^{(i)}\rangle+\sqrt{1-r_{i}}|\Phi_{abp}^{(i)}\rangle,\hskip
3mm (i=1,2),
\end{equation}
then there are corresponding unitary operators $U_{B}$ and
$U_{A}$:
\begin{equation}
U_{B}(|\phi_{i}\rangle|\Sigma\rangle|P_{0}\rangle)=\sqrt{r_{i}^{B}}|\psi_{i}\rangle^{\otimes
m}|P^{(i)}_{B}\rangle+\sqrt{1-r_{i}^{B}}|\Phi_{abp_{B}}^{(i)}\rangle,\hskip
3mm (i=1,2),
\end{equation}
\begin{equation}
U_{A}(|\psi_{i}\rangle|\Sigma\rangle|P_{0}\rangle)=\sqrt{r_{i}^{A}}|\psi_{i}\rangle^{\otimes
(m+1)}|P^{(i)}_{A}\rangle+\sqrt{1-r_{i}^{B}}|\Phi_{abp_{A}}^{(i)}\rangle,\hskip
3mm (i=1,2),
\end{equation}
such that $r_{i}^{B}+(1-r_{i}^{B})r_{i}^{A}\geq r_{i}$ $(i=1,2)$,
where $r_{i}$, $r_{i}^{B}$, and $r_{i}^{A}$ denote the success
probabilities in the three machines, respectively, and  $\langle
P^{(i)}|\Phi_{abp}^{(j)}\rangle=\langle
P^{(i)}_{B}|\Phi_{abp_{B}}^{(j)}\rangle=\langle
P^{(i)}_{A}|\Phi_{abp_{A}}^{(j)}\rangle=0$ for any $i,j\in
\{1,2\}$. The above implication means that when the state chosen
from two nonorthogonal states, the best efficiency of producing
$m+1$ copies is always achieved by a two-step cloning protocol in
which the auxiliary party first tries to produce $m$ copies from
the supplementary state, and if it fails, then the original state
is used to produce $m+1$ copies by means of the probabilistic
cloning device proposed by Duan and Guo [29,30]. For the sake of
simplicity, we may represent the cloning devices described by Eqs.
(13,14,15) as:
\begin{equation}
|\psi_{i}\rangle|\phi_{i}\rangle\stackrel{r_{i}}{\longrightarrow}|\psi_{i}\rangle^{m+1},
\hskip 3mm (i=1,2),
\end{equation}
\begin{equation}
\Longrightarrow
|\phi_{i}\rangle\stackrel{r_{i}^{B}}{\longrightarrow}|\psi_{i}\rangle^{m}\hskip
2mm {\rm and}\hskip 2mm
|\psi_{i}\rangle\stackrel{r_{i}^{A}}{\longrightarrow}|\psi_{i}\rangle^{m+1},
\hskip 3mm (i=1,2).
\end{equation}
However, when the state chosen from $n$ states, with $n>2$ and
without orthogonal pairs of states, the above implication
described by Eqs. (16,17) may not hold again, i.e., the best
efficiency is not always reached by such a two-step cloning
protocol [38].

In this paper, we will show the following equivalent relation: For
any two non-orthogonal states $|\psi_{1}\rangle$ and
$|\psi_{2}\rangle$, and supplementary states $|\phi_{1}\rangle$
and $|\phi_{2}\rangle$, there exists unitary operator $U$:
\begin{equation}
U(|\psi_{i}\rangle|\phi_{i}\rangle|P_{0}\rangle)=\sum_{k=1}^{m}\sqrt{r_{k}^{(i)}}|\psi_{i}\rangle^{\otimes(k+1)}|0\rangle^{\otimes(m-k)}|P_{k}^{(i)}\rangle
+\sum_{l=m+1}^{N}\sqrt{f_{l}^{(i)}}|\Psi_{l}\rangle_{AB}|P_{l}\rangle,
\hskip 3mm (i=1,2),
\end{equation}
where $|P_{1}^{(i)}\rangle$, $|P_{2}^{(i)}\rangle$, $\ldots$,
$|P_{m}^{(i)}\rangle$, $|P_{m+1}\rangle$, $|P_{m+2}\rangle$,
$\ldots$, $|P_{N}\rangle$ are orthonormal for any $i\in\{1,2\}$,
if and only if there are unitary operators $U_{B}$ and $U_{A}$:
\begin{equation}
U_{B}(|\phi_{i}\rangle|\Sigma\rangle|P_{0}\rangle)=\sum_{k=1}^{m}\sqrt{r_{k,B}^{(i)}}|\psi_{i}\rangle^{\otimes(k)}|0\rangle^{\otimes(m-k+1)}|P_{k,B}^{(i)}\rangle
+\sum_{l=m+1}^{N}\sqrt{f_{l,B}^{(i)}}|\Phi_{l}^{(B)}\rangle_{AB}|P_{l,B}\rangle,\hskip
3mm (i=1,2),
\end{equation}
\begin{equation}
U_{A}(|\psi_{i}\rangle|\Sigma\rangle|P_{0}\rangle)=\sum_{k=1}^{m}\sqrt{r_{k,A}^{(i)}}|\psi_{i}\rangle^{\otimes(k+1)}|0\rangle^{\otimes(m-k)}|P_{k,A}^{(i)}\rangle
+\sum_{l=m+1}^{N}\sqrt{f_{l,A}^{(i)}}|\Phi_{l}^{(A)}\rangle_{AB}|P_{l,A}\rangle,\hskip
3mm (i=1,2),
\end{equation}
where $|P_{1,B}^{(i)}\rangle$, $|P_{2,B}^{(i)}\rangle$, $\ldots$,
$|P_{m,B}^{(i)}\rangle$, $|P_{m+1,B}\rangle$, $|P_{m+2,B}\rangle$,
$\ldots$, $|P_{N,B}\rangle$ are orthonormal, and, also,
$|P_{1,A}^{(i)}\rangle$, $|P_{2,A}^{(i)}\rangle$, $\ldots$,
$|P_{m,A}^{(i)}\rangle$, $|P_{m+1,A}\rangle$, $|P_{m+2,A}\rangle$,
$\ldots$, $|P_{N,A}\rangle$ are orthonormal for any $i\in\{1,2\}$;
$r_{k}^{(i)}$, $r_{k,B}^{(i)}$, and $r_{k,A}^{(i)}$ represent the
success probabilities for producing
$|\psi_{i}\rangle^{\otimes(k+1)}$, $|\psi_{i}\rangle^{\otimes k}$,
and $|\psi_{i}\rangle^{\otimes(k+1)}$, respectively, in three
cloning devices.

Furthermore, it is satisfied that if the unitary transformation
described by Eq. (18) holds, then there exist unitary
transformations described by Eqs. (19,20) such that
\begin{equation}
\sum_{k=1}^{m}r_{k,B}^{(i)}+\left(1-\sum_{k=1}^{m}r_{k,B}^{(i)}\right)\sum_{k=1}^{m}r_{k,A}^{(i)}\geq
\sum_{k=1}^{m}r_{i}^{(k)}, \hskip 3mm (i=1,2);
\end{equation}
conversely, if Eqs. (19,20) hold, then there is unitary
transformation by Eq. (18) satisfying
\begin{equation}
\sum_{k=1}^{m}r_{k,B}^{(i)}+\left(1-\sum_{k=1}^{m}r_{k,B}^{(i)}\right)\sum_{k=1}^{m}r_{k,A}^{(i)}\leq
\sum_{k=1}^{m}r_{i}^{(k)},\hskip 3mm (i=1,2).
\end{equation}
In the interest of simplicity, we may represent the above Eqs.
(18,19,20) as:
\begin{equation}
|\psi_{i}\rangle|\phi_{i}\rangle\stackrel{\sum_{k=1}^{m}r_{k}^{(i)}}{\longrightarrow}\sum_{k=1}^{m}|\psi_{i}\rangle^{\otimes(k+1)},\hskip
3mm (i=1,2),
\end{equation}
 \hskip 40mm $\Longleftrightarrow$
\begin{equation}
|\phi_{i}\rangle\stackrel{\sum_{k=1}^{m}r_{k,B}^{(i)}}{\longrightarrow}\sum_{k=1}^{m}|\psi_{i}\rangle^{\otimes(k)}
\end{equation}
and
\begin{equation}
|\psi_{i}\rangle\stackrel{\sum_{k=1}^{m}r_{k,A}^{(i)}}{\longrightarrow}\sum_{k=1}^{m}|\psi_{i}\rangle^{\otimes(k+1)},\hskip
3mm (i=1,2).
\end{equation}
Note that transformation (20) is exactly the NCM studied by Pati
[13] and stated above. The above equivalence shows that the
optimal efficiency of the NCMSI in which the original party and
the supplementary party can perform quantum communication equals
the optimal efficiency achieved by the two-step cloning protocol
wherein classical communication is only allowed between the
original and the supplementary parties. Therefore, in regard to
the optimal success probabilities, if
$\sum_{k=1}^{m}r_{k,B}^{(i)}>0$, then
$\sum_{k=1}^{m}r_{k}^{(i)}>\sum_{k=1}^{m}r_{k,A}^{(i)}$,
$(i=1,2)$, which implies that the NCMSI may increase the success
probability. As well, if we take only one $r_{k,B}^{(i)}$ and one
$r_{k,A}^{(i)}$ nonzero for some $k$, then our right-implication
reduces to the implication described by transformations (16,17).
Therefore, our result generalizes and completes the result proved
by Azuma {\it et al.} [38].

\section*{3. Proofs of main results}
\hskip 6mm Firstly, for the sake of readability, we still quickly
review the results by Azuma {\it et al.} [38], and present some
transformations, some of which were indeed described before.

Probabilistic cloning machine firstly posed by Duan and Guo
[29,30] describes that for any state set
$\{|\psi_{1}\rangle,|\psi_{2}\rangle,\ldots,|\psi_{k}\rangle\}$,
there exists unitary operator $U$ such that
\begin{equation}
U(|\psi_{i}\rangle|\Sigma\rangle|P_{0}\rangle)=\sqrt{r_{i}}|\psi_{i}\rangle|\psi_{i}\rangle|P^{(i)}\rangle+\sqrt{1-r_{i}}|\Phi_{ABP}^{(i)}\rangle,
\hskip 3mm (i=1,2,\ldots,k),
\end{equation}
if and only if matrix
$X^{(1)}-\sqrt{\Gamma}X^{(2)}\sqrt{\Gamma^{\dag}}$ is positive
semidefinite, where $X^{(1)}=[\langle\psi_{i}|\psi_{j}\rangle]$,
$X^{(2)}=[\langle\psi_{i}|\psi_{j}\rangle^{2}\langle
P^{(i)}|P^{(j)}\rangle]$, $\sqrt{\Gamma}=\sqrt{\Gamma^{\dag}}={\rm
diag}(\sqrt{r_{1}},\sqrt{r_{2}},\ldots,\sqrt{r_{k}})$. The
efficiency of cloning is as $\sum_{i=1}^{k}p_{i}r_{i}$ if $p_{i}$
are the probabilities for choosing states $|\psi_{i}\rangle$
$(i=1,2,\ldots,k)$.

Azuma {\it et al.} [38] showed that for two nonorthogonal states,
$|\psi_{i}\rangle$ $(i=1,2)$, if there exists unitary operator $U:
|\psi_{i}\rangle|\phi_{i}\rangle\rightarrow
\sqrt{r_{i}}|\psi_{i}\rangle^{\otimes (m+1)}$ (for simplicity,
they left out the failure item and the states of the probe
device), then there also exist unitary operator
$U_{A}:|\psi_{i}\rangle\rightarrow
\sqrt{r_{i}^{A}}|\psi_{i}\rangle^{\otimes (m+1)}$ and unitary
operator $U_{B}:|\phi_{i}\rangle\rightarrow
\sqrt{r_{i}^{B}}|\psi_{i}\rangle^{\otimes (m)}$ satisfying
$r_{i}^{B}+(1-r_{i}^{B})r_{i}^{A}\geq r_{i}$ $(i=1,2)$. For $k$
states with $k\geq 3$, they verified that there exist state sets
$\{|\psi_{i}\rangle\}$ and $\{|\phi_{i}\rangle\}$, as well as
unitary operator $U$ above, such that for any unitary operators
$U_{A}$ and $U_{B}$ above, it holds that $r_{i}^{A}=0$
$(i=1,2,\ldots,n)$, and
$\sum_{i=1}^{k}\frac{1}{n}r_{i}>\sum_{i=1}^{k}\frac{1}{n}r_{i}^{B}$.

We enter on our discussion. Suppose Alice holds the original copy
$|\psi_{i}\rangle$ and Bob possesses the supplementary information
$|\phi_{i}\rangle$ $(i=1,2)$. If Alice and Bob are allowed to
communicate with one-way quantum channel from Bob to Alice, then a
single party holding both the original and the supplementary
information $|\psi_{i}\rangle|\phi_{i}\rangle$ performs the
following cloning process described by a unitary operator $U$:
\begin{equation}
U(|\psi_{i}\rangle|\phi_{i}\rangle|P_{0}\rangle)=\sum_{k=1}^{m}\sqrt{r_{k}^{(i)}}|\psi_{i}\rangle^{\otimes(k+1)}|0\rangle^{\otimes(m-k)}|P_{k}^{(i)}\rangle
+\sum_{l=m+1}^{N}\sqrt{f_{l}^{(i)}}|\Psi_{l}\rangle_{AB}|P_{l}\rangle,\hskip
3mm (i=1,2),
\end{equation}
where $0\leq r_{k}^{(i)}\leq 1$ for $k=1,2,\ldots,m$, and
$\sum_{k=1}^{m}r_{k}^{(i)}<1$ (in terms of [13],
$\sum_{k=1}^{m}r_{k}^{(i)}=1$ is impossible), $|P_{0}\rangle$,
$|P_{k}^{(i)}\rangle$, and $|P_{l}\rangle$ are the states of the
probing device, satisfying that $|P_{1}^{(i)}\rangle$,
$|P_{2}^{(i)}\rangle$, $\ldots$, $|P_{m}^{(i)}\rangle$,
$|P_{m+1}\rangle$, $|P_{m+2}\rangle$, $\ldots$, $|P_{N}\rangle$
are orthonormal for $i=1,2$. Moreover, $N>m$, $|0\rangle$ is the
state of the ancillary system $B$, $r_{k}^{(i)}$ and $f_{l}^{(i)}$
are the success and the failure probabilities, respectively. If
$p_{i}$ are {\it a priori} probabilities for choosing
$|\psi_{i}\rangle|\phi_{i}\rangle$ ($i=1,2$), then the global
success probability $P_{s}$ for copying is
\begin{equation}
P_{s}=\sum_{i=1}^{2}p_{i}\sum_{k=1}^{m}r_{k}^{(i)}.
\end{equation}
If Alice and Bob only can use classical channel for communication,
they may respectively run the following machines described by
unitary operators $U_{A}$ and $U_{B}$, where $U_{A}$ is exactly
Pati's NCM [13]:
\begin{equation}
U_{A}(|\psi_{i}\rangle|\Sigma\rangle|P_{0}\rangle)=\sum_{k=1}^{m}\sqrt{r_{k,A}^{(i)}}|\psi_{i}\rangle^{\otimes(k+1)}|0\rangle^{\otimes(m-k)}|P_{k,A}^{(i)}\rangle
+\sum_{l=m+1}^{N}\sqrt{f_{l,A}^{(i)}}|\Phi_{l}^{(A)}\rangle_{AB}|P_{l,A}\rangle,
\hskip 3mm (i=1,2),
\end{equation}
such that $0\leq r_{k,A}^{(i)}\leq 1$ for $k=1,2,\ldots,m$, where
$|P_{0}\rangle$, $|P_{k,A}^{(i)}\rangle$, and $|P_{l,A}\rangle$
are the states of the probe device, satisfying that
$|P_{1,A}^{(i)}\rangle$, $|P_{2,A}^{(i)}\rangle$, $\ldots$,
$|P_{m,A}^{(i)}\rangle$, $|P_{m+1,A}\rangle$, $|P_{m+2,A}\rangle$,
$\ldots$, $|P_{N,A}\rangle$ are orthonormal for $i=1,2$. If
$p_{i}^{(A)}$ are {\it a priori} probabilities for choosing
$|\psi_{i}\rangle$ ($i=1,2$), then the global success probability
$P_{s}^{(A)}$ for copying is
\begin{equation}
P_{s}^{(A)}=\sum_{i=1}^{2}p_{i}^{(A)}\sum_{k=1}^{m}r_{k,A}^{(i)}.
\end{equation}
$U_{B}$ is as follows:
\begin{equation}
U_{B}(|\phi_{i}\rangle|\Sigma\rangle|P_{0}\rangle)=\sum_{k=1}^{m}\sqrt{r_{k,B}^{(i)}}|\psi_{i}\rangle^{\otimes
k}|0\rangle^{\otimes(m-k+1)}|P_{k,B}^{(i)}\rangle
+\sum_{l=m+1}^{N}\sqrt{f_{l,B}^{(i)}}|\Phi_{l}^{(B)}\rangle_{AB}|P_{l,B}\rangle,\hskip
3mm (i=1,2),
\end{equation}
such that $0\leq r_{k,B}^{(i)}\leq 1$ for $k=1,2,\ldots,m$, where
$|P_{0}\rangle$, $|P_{k,B}^{(i)}\rangle$, and $|P_{l,B}\rangle$
are the states of the probe device, satisfying that
$|P_{1,B}^{(i)}\rangle$, $|P_{2,B}^{(i)}\rangle$, $\ldots$,
$|P_{m,B}^{(i)}\rangle$, $|P_{m+1,B}\rangle$, $|P_{m+2,B}\rangle$,
$\ldots$, $|P_{N,B}\rangle$ are orthonormal for $i=1,2$. If
$p_{i}^{(B)}$ are {\it a priori} probabilities for choosing
$|\psi_{i}\rangle$ ($i=1,2$), then the global success probability
$P_{s}^{(B)}$ for copying is
\begin{equation}
P_{s}^{(B)}=\sum_{i=1}^{2}p_{i}^{(B)}\sum_{k=1}^{m}r_{k,B }^{(i)}.
\end{equation}
If Alice and Bob only can use {\it one-way} classical channel for
communication from {\it Bob to Alice}, then Bob first performs
machine described by  Eq. (31), and tells Alice the result of
success or failure. If Bob succeeds, Alice only preserves her copy
as is; otherwise, Alice runs the machine described by Eq. (29).
Therefore, in this case, the success probability for producing
quantum superposition of multiple clones
$\sum_{k=1}^{m}|\psi_{i}\rangle^{\otimes (k+1)}$ when inputting
$|\psi_{i}\rangle|\phi_{i}\rangle$, is
\begin{equation}
\sum_{l=1}^{m}r_{k,B}^{(i)}+\left(1-\sum_{l=1}^{m}r_{l,B}^{(i)}\right)\sum_{l=1}^{m}r_{k,A}^{(i)}.
\end{equation}
Similarly, if Alice and Bob can use only {\it one-way} classical
channel for communication from {\it Alice to Bob}, then Alice
first performs Pati's machine described by  Eq. (29), and then
tells Bob the result of success or failure. If Alice succeeds, Bob
does nothing; otherwise, Bob runs machine by Eq. (31). Thus, it is
seen that the success probability for producing quantum
superposition of multiple clones
$\sum_{k=1}^{m}|\psi_{i}\rangle^{\otimes (k+1)}$ with input
$|\psi_{i}\rangle|\phi_{i}\rangle$ is
\begin{equation}
\sum_{k=1}^{m}r_{k,A}^{(i)}+\left(1-\sum_{l=1}^{m}r_{l,A}^{(i)}\right)r_{k,B}^{(i)}.
\end{equation}
If Alice and Bob can use {\it two-way} classical channel for
communication, i.e., they can communicate each other, then they
first independently carry out machines described by Eqs. (29,31),
and, afterwards, inform the other of the outcome produced.
Therefore, the success probability for producing quantum
superposition of multiple clones
$\sum_{k=1}^{m}|\psi_{i}\rangle^{\otimes (k+1)}$ with input
$|\psi_{i}\rangle|\phi_{i}\rangle$ will be
\begin{equation}1-\left(1-\sum_{k=1}^{m}r_{k,A}^{(i)}\right)\left(1-\sum_{k=1}^{m}r_{k,B}^{(i)}\right)
=\sum_{k=1}^{m}r_{k,A}^{(i)}+\sum_{k=1}^{m}r_{k,B}^{(i)}-\sum_{k=1}^{m}r_{k,A}^{(i)}\sum_{k=1}^{m}r_{k,B}^{(i)}.
\end{equation}
Notably, whichever classical communication we choose, it is
clearly seen that with input $|\psi_{i}\rangle|\phi_{i}\rangle$,
the success probabilities for producing quantum superposition of
multiple clones $\sum_{k=1}^{m+1}|\psi_{i}\rangle^{\otimes (k+1)}$
are equal.

In what follows, we denote
$\alpha=\langle\psi_{1}|\psi_{2}\rangle$,
$\beta=\langle\phi_{1}|\phi_{2}\rangle$, $p_{k}=\langle
P_{k}^{(1)}|P_{k}^{(2)}\rangle$, $p_{k,A}=\langle
P_{k,A}^{(1)}|P_{k,A}^{(2)}\rangle$, $p_{k,B}=\langle
P_{k,B}^{(1)}|P_{k,B}^{(2)}\rangle$. Now we notice that Eqs.
(27,29,31), hold if and only if the matrices

$Z^{(1)}-\sum_{k=1}^{m}\sqrt{\Gamma_{k}}G^{(m+1)}\sqrt{\Gamma_{k}^{\dag}}$,

$X^{(1)}-\sum_{k=1}^{m}\sqrt{\Gamma_{k,A}}G_{A}^{(m+1)}\sqrt{\Gamma_{k,A}^{\dag}}$,

$Y^{(1)}-\sum_{k=1}^{m}\sqrt{\Gamma_{k,B}}G_{B}^{(m+1)}\sqrt{\Gamma_{k,B}^{\dag}}$,\\
are positive semidefinite, respectively, where
$Z^{(1)}=[\langle\psi_{i}|\psi_{j}\rangle\langle\phi_{i}|\phi_{j}\rangle]$,
$X^{(1)}=[\langle\psi_{i}|\psi_{j}\rangle]$, and
$Y^{(1)}=[\langle\phi_{i}|\phi_{j}\rangle]$;
$G^{(m+1)}=[\langle\psi_{i}|\psi_{j}\rangle^{m+1}\langle
P_{k}^{(i)}|P_{k}^{(j)}\rangle],$
$G_{A}^{(m+1)}=[\langle\psi_{i}|\psi_{j}\rangle^{m+1}\langle
P_{k,A}^{(i)}|P_{k,A}^{(j)}\rangle]$, and\\
 $G_{B}^{(m)}=[\langle\psi_{i}|\psi_{j}\rangle^{m}\langle
P_{k,B}^{(i)}|P_{k,B}^{(j)}\rangle]$; $\sqrt{\Gamma_{k}}={\rm
diag}(r_{k}^{(1)},r_{k}^{(2)})$, $\sqrt{\Gamma_{k,A}}={\rm
diag}(r_{k,A}^{(1)},r_{k,A}^{(2)})$, and $\sqrt{\Gamma_{k,B}}={\rm
diag}(r_{k,B}^{(1)},r_{k,B}^{(2)})$. Furthermore, we note that the
three matrices above are positive semidefinite if and only if
their determinants are nonnegative, respectively, that is,
\begin{equation}
\sqrt{(1-\sum_{k=1}^{m}r_{k}^{(1)})(1-\sum_{k=1}^{m}r_{k}^{(2)})}-|\alpha\beta-\sum_{k=1}^{m}\sqrt{r_{k}^{(1)}r_{k}^{(2)}}\alpha^{k+1}p_{k}|\geq
0,
\end{equation}
\begin{equation}
\sqrt{(1-\sum_{k=1}^{m}r_{k,A}^{(1)})(1-\sum_{k=1}^{m}r_{k,A}^{(2)})}-|\alpha-\sum_{k=1}^{m}\sqrt{r_{k,A}^{(1)}r_{k,A}^{(2)}}\alpha^{k+1}p_{k,A}|\geq
0,
\end{equation}
\begin{equation}
\sqrt{(1-\sum_{k=1}^{m}r_{k,B}^{(1)})(1-\sum_{k=1}^{m}r_{k,B}^{(2)})}-|\beta-\sum_{k=1}^{m}\sqrt{r_{k,B}^{(1)}r_{k,B}^{(2)}}\alpha^{k}p_{k,B}|\geq
0.
\end{equation}
If $|\beta|>
\sum_{k=1}^{m}\sqrt{r_{k}^{(1)}r_{k}^{(2)}}|\alpha|^{k}$, then, by
taking appropriate amplitudes of $p_{k}$, Ineq. (36) is equivalent
to
\begin{equation}
\sqrt{(1-\sum_{k=1}^{m}r_{k}^{(1)})(1-\sum_{k=1}^{m}r_{k}^{(2)})}-|\alpha\beta|+\sum_{k=1}^{m}\sqrt{r_{k}^{(1)}r_{k}^{(2)}}|\alpha|^{k+1}\geq
0;
\end{equation}
analogously, if $1>
\sum_{k=1}^{m}\sqrt{r_{k,A}^{(1)}r_{k,A}^{(2)}}|\alpha|^{k}$ and
$|\beta|>
\sum_{k=1}^{m}\sqrt{r_{k,B}^{(1)}r_{k,B}^{(2)}}|\alpha|^{k}$ hold,
respectively, then correspondingly, Ineqs. (38,39) are
respectively equivalent to
\begin{equation}
\sqrt{(1-\sum_{k=1}^{m}r_{k,A}^{(1)})(1-\sum_{k=1}^{m}r_{k,A}^{(2)})}-|\alpha|+\sum_{k=1}^{m}\sqrt{r_{k,A}^{(1)}r_{k,A}^{(2)}}|\alpha|^{k+1}\geq
0,
\end{equation}
\begin{equation}
\sqrt{(1-\sum_{k=1}^{m}r_{k,B}^{(1)})(1-\sum_{k=1}^{m}r_{k,B}^{(2)})}-|\beta|+\sum_{k=1}^{m}\sqrt{r_{k,B}^{(1)}r_{k,B}^{(2)}}|\alpha|^{k}\geq
0.
\end{equation}

With input $|\psi_{i}\rangle|\phi_{i}\rangle$, the efficiency of
producing quantum superposition of multiple clones
$\sum_{k=1}^{m}|\psi_{i}\rangle^{\otimes (k+1)}$ which Alice and
Bob achieve via quantum channel can always be achieved by a
two-step cloning protocol in which Alice and Bob are only allowed
to execute {\it one-way} or {\it two-way} classical communication.
This is described by the following Theorem 1.

 {\it Theorem 1.}--If there exists unitary operator $U$ such that
Eq. (27) holds, then there are unitary operators $U_{A}$ and
$U_{B}$ satisfying Eqs. (29,31), respectively, such that
\begin{equation}\sum_{k=1}^{m}r_{k}^{(i)}\leq
\sum_{k=1}^{m}r_{k,B}^{(i)}+(1-\sum_{k=1}^{m}r_{k,B}^{(i)})\sum_{k=1}^{m}r_{k,A}^{(i)},\end{equation}
for $i=1,2$.

{\it Proof:}  As above, denote
$\alpha=\langle\psi_{1}|\psi_{2}\rangle$,
$\beta=\langle\phi_{1}|\phi_{2}\rangle$.

{\it Case 1.}
$|\beta|\leq\sum_{k=1}^{m}\sqrt{r_{k}^{(1)}r_{k}^{(2)}}|\alpha|^{k}$.
In this case, we only take any $r_{k,B}^{(i)}$ satisfying
$r_{k,B}^{(i)}\geq r_{k}^{(i)}$ for $k=1,2,\ldots,m$, and
$\sum_{k=1}^{m}r_{k,B}^{(i)}=1$ ($i=1,2$). Clearly,
$|\beta|\leq\sum_{k=1}^{m}\sqrt{r_{k,B}^{(1)}r_{k,B}^{(2)}}|\alpha|^{k}$
also holds. Then it suffices to take appropriate $p_{k,B}$ such
that
$\beta-\sum_{k=1}^{m}\sqrt{r_{k,B}^{(1)}r_{k,B}^{(2)}}\alpha^{k}p_{k,B}=0$.
Thus, Ineq. (38) holds. By taking $r_{k,A}^{(i)}=0$ $(1\leq k\leq
m, 1\leq i\leq 2)$, then Ineq. (37) holds. So, the theorem is
proved in this situation.

{\it Case 2.}
$|\beta|>\sum_{k=1}^{m}\sqrt{r_{k}^{(1)}r_{k}^{(2)}}|\alpha|^{k}$.
We set a function $F$ from $[0,1]^{m}\times [0,1]^{m}$ to
$[0,+\infty)$ as:
\begin{equation}
F(x_{1},x_{2},\ldots,x_{m};y_{1},y_{2},\ldots,y_{m})=\frac{\sqrt{(1-\sum_{k=1}^{m}x_{k})(1-\sum_{k=1}^{m}y_{k})}}
{|\beta|-\sum_{k=1}^{m}\sqrt{x_{k}y_{k}}|\alpha|^{k}}.
\end{equation}
Clearly function $F$ is continuous on $[0,1]^{m}\times [0,1]^{m}$,
and
\begin{equation}
F(0,0,\ldots,0;0,0,\ldots,0)=\frac{1}{|\beta|}\geq 1,
\end{equation}
as well as, by Ineq. (39),
\begin{equation}
F(r_{1}^{(1)},r_{2}^{(1)},\ldots,r_{m}^{(1)};r_{1}^{(2)},r_{2}^{(2)},\ldots,r_{m}^{(2)})\geq
|\alpha|.
\end{equation}
To prove the theorem, we somewhat change function $F$ to set up a
new function $H$ that  only has $m$ variables at most. The main
idea to establish $H$ is to reduce the number $2m$ of the
variables in $F$ to not more than $m$, and we present the way of
constructing function $H$ from function $F$ in detail:

(i) For $1\leq k\leq m$, if $0\not=r_{k}^{(1)}\geq r_{k}^{(2)}$,
then the pair of variables $(x_{k},y_{k})$ in $F$ will be replaced
by $(x_{k},c_{k}x_{k})$, where
$\frac{r_{k}^{(2)}}{r_{k}^{(1)}}=c_{k}\leq 1$; if
$0=r_{k}^{(1)}\geq r_{k}^{(2)}$, then the pair of variables
$(x_{k},y_{k})$ in $F$ will be replaced by the pair $(0,0)$ of
constants.

(ii) For $1\leq k\leq m$, if $r_{k}^{(1)}<r_{k}^{(2)}$, we replace
the pair of variables $(x_{k},y_{k})$ in $F$  by
$(c_{k}^{'}y_{k},y_{k})$, where
$c_{k}^{'}=\frac{r_{k}^{(1)}}{r_{k}^{(2)}}\leq 1$.

By means of the above way to adjust and decrease those variables
in function $F$, we obtain a new function $H$ whose number of
variables is at most $m$, instead of $2m$, that is the form: For
$z_{k}\in \{x_{k},y_{k}\}$, $1\leq k\leq m$,
\begin{equation}
H(z_{1},z_{2},\ldots,z_{m})=F(u_{1},u_{2},\ldots,u_{m};v_{1},v_{2},\ldots,v_{m}),
\end{equation}
where:

(i) If $0\not=r_{k}^{(1)}\geq r_{k}^{(2)}$, then $z_{k}=x_{k}$,
and, $u_{k}=x_{k}$, $v_{k}=c_{k}x_{k}$, where
$c_{k}=\frac{r_{k}^{(2)}}{r_{k}^{(1)}}\leq 1$.

(ii) If $0=r_{k}^{(1)}\geq r_{k}^{(2)}$, then
$z_{k}=u_{k}=v_{k}=0$.

(iii) If $\frac{r_{k}^{(1)}}{r_{k}^{(2)}}< 1$, then $z_{k}=y_{k}$,
and, $u_{k}=c_{k}^{'}y_{k}$, $v_{k}=y_{k}$, where
$c_{k}^{'}=\frac{r_{k}^{(1)}}{r_{k}^{(2)}}$.

Without loss of generality, we suppose that always
$r_{k}^{(1)}\geq r_{k}^{(2)}$, $k=1,2,\ldots,m$. Then we have
\begin{equation}
H(x_{1},x_{2},\ldots,x_{m})=F(x_{1},x_{2},\ldots,x_{m};c_{1}x_{1},c_{2}x_{2},\ldots,c_{m}x_{m}),
\end{equation}
where when $r_{k}^{(1)}=0$, $x_{k}\equiv 0$, $(k=1,2,\ldots,m)$.

 By Ineqs. (44,45),
\begin{eqnarray}
&&\nonumber H(0,0,\ldots,0)\\
&=&F(0,0,\ldots,0;0,0,\ldots,0)\\
&=&\frac{1}{|\beta|}\geq 1,
\end{eqnarray}
and
\begin{eqnarray}
&&\nonumber H(r_{1}^{(1)},r_{2}^{(1)},\ldots,r_{m}^{(1)})\\
&=&F(r_{1}^{(1)},r_{2}^{(1)},\ldots,r_{m}^{(1)};r_{1}^{(2)},r_{2}^{(2)},\ldots,r_{m}^{(2)})\\
&\geq& |\alpha|.
\end{eqnarray}
Next we consider two scenarios to complete the proof:

(I) If $H(r_{1}^{(1)},r_{2}^{(1)},\ldots,r_{m}^{(1)})\geq 1$, then
\begin{eqnarray}
&&\nonumber F(r_{1}^{(1)},r_{2}^{(1)},\ldots,r_{m}^{(1)};r_{1}^{(2)},r_{2}^{(2)},\ldots,r_{m}^{(2)})\\
&=&H(r_{1}^{(1)},r_{2}^{(1)},\ldots,r_{m}^{(1)})\\
&\geq& 1,
\end{eqnarray}
and, therefore, by Eq. (43) we have
\begin{eqnarray}
&&\nonumber F(r_{1}^{(1)},r_{2}^{(1)},\ldots,r_{m}^{(1)};r_{1}^{(2)},r_{2}^{(2)},\ldots,r_{m}^{(2)})\\
&=&\nonumber
\frac{\sqrt{(1-\sum_{k=1}^{m}r_{k}^{(1)})(1-\sum_{k=1}^{m}r_{k}^{(2)})}}
{|\beta|-\sum_{k=1}^{m}\sqrt{r_{k}^{(1)}r_{k}^{(2)}}|\alpha|^{k}}\\
&\geq& 1.
\end{eqnarray}
Therefore, by taking $r_{k,B}^{(i)}=r_{k}^{(i)}$,
$(k=1,2,\ldots,m;i=1,2)$, Ineq. (41) holds. As a result, there
exist unitary operators $U_{A}$ and $U_{B}$ such that Eqs. (29,31)
hold, in which we can chose $r_{k,A}^{(i)}=0$ and
$r_{k,B}^{(i)}=r_{k}^{(i)}$, $(k=1,2,\ldots,m;i=1,2)$. In this
case, the theorem is proved.

(II) If $H(r_{1}^{(1)},r_{2}^{(1)},\ldots,r_{m}^{(1)})< 1$, then,
together with $H(0,0,\ldots,0)\geq 1$ (i.e., Eq. (49)), by {\it
intermediate value theorem of continuous functions}, there exist
$r_{k,B}^{(1)}$ such that \begin{equation} 0\leq r_{k,B}^{(1)}\leq
r_{k}^{(1)},\hskip 5mm (k=1,2,\ldots,m),
\end{equation}
and
\begin{equation}
H(r_{1,B}^{(1)},r_{2,B}^{(1)},\ldots,r_{m,B}^{(1)})=1.
\end{equation}
Now, for $k=1,2,\ldots,m$, we take
\begin{equation}
r_{k,B}^{(2)}=\left\{\begin{array}{ll} 0,& {\rm if}\hskip 2mm
r_{k}^{(1)}=0,\\
\frac{r_{k}^{(2)}}{r_{k}^{(1)}}r_{k,B}^{(1)},& {\rm otherwise}.
\end{array}
\right.
\end{equation}
Denoting $c_{k}=\left\{\begin{array}{ll} 0,& {\rm if}\hskip 2mm
r_{k}^{(1)}=0,\\
\frac{r_{k}^{(2)}}{r_{k}^{(1)}},&{\rm otherwise},
\end{array}\right.$ then clearly we have
\begin{equation}
r_{k,B}^{(2)}=c_{k}r_{k,B}^{(1)}, \hskip 5mm
r_{k}^{(2)}=c_{k}r_{k}^{(1)},
\end{equation}
for $k=1,2,\ldots,m$ and $i=1,2$; as well, by Ineqs. (55,58),
$\sum_{k=1}^{m}r_{k,B}^{(i)}\leq \sum_{k=1}^{m}r_{k}^{(i)}$ holds
for $i=1,2$. Now we take
\begin{equation}
r_{k,A}^{(i)}=\frac{r_{k}^{(i)}-r_{k,B}^{(i)}}{1-\sum_{k=1}^{m}r_{k,B}^{(i)}},
\hskip 5mm (i=1,2),
\end{equation}
then
\begin{eqnarray}
\nonumber \sqrt{(1-r_{k,A}^{(1)})(1-r_{k,A}^{(2)})}&=&
\sqrt{\frac{(1-\sum_{k=1}^{m}r_{k}^{(1)})(1-\sum_{k=1}^{m}r_{k}^{(2)})}{(1-\sum_{k=1}^{m}r_{k,B}^{(1)})(1-\sum_{k=1}^{m}r_{k,B}^{(2)})}}\\
&\geq&\frac{|\beta|-\sum_{k=1}^{m}\sqrt{r_{k}^{(1)}r_{k}^{(2)}}|\alpha|^{k}}{|\beta|-\sum_{k=1}^{m}\sqrt{r_{k,B}^{(1)}r_{k,B}^{(2)}}|\alpha|^{k}}|\alpha|,
\end{eqnarray}
and
\begin{equation}
\sqrt{r_{k,A}^{(1)}r_{k,A}^{(2)}}=
\frac{(r_{k}^{(1)}-r_{k,B}^{(1)})(r_{k}^{(2)}-r_{k,B}^{(2)})}{|\beta|-\sum_{k=1}^{m}\sqrt{r_{k,B}^{(1)}r_{k,B}^{(2)}}|\alpha|^{k}}.
\end{equation}
By Ineq. (60) and Eq. (61) we have
\begin{eqnarray}
\nonumber &&\sqrt{(1-\sum_{k=1}^{m}r_{k,A}^{(1)})(1-\sum_{k=1}^{m}r_{k,A}^{(2)})}-|\alpha|+\sum_{k=1}^{m}\sqrt{r_{k,A}^{(1)}r_{k,A}^{(2)}}|\alpha|^{k+1}\\
&\geq&
\frac{\sum_{k=1}^{m}\left(\sqrt{r_{k,B}^{(1)}r_{k,B}^{(2)}}-\sqrt{r_{k}^{(1)}r_{k}^{(2)}}+\sqrt{(r_{k}^{(1)}-r_{k,B}^{(1)})(r_{k}^{(2)}-r_{k,B}^{(2)})}\right)|\alpha|^{k+1}}
{|\beta|-\sum_{k=1}^{m}\sqrt{r_{k,B}^{(1)}r_{k,B}^{(2)}}}.
\end{eqnarray}
Due to Eq. (58), i.e., $r_{k,B}^{(2)}=c_{k}r_{k,B}^{(1)}, \hskip
2mm r_{k}^{(2)}=c_{k}r_{k}^{(1)}$, we have
\begin{equation}
\sqrt{(r_{k}^{(1)}-r_{k,B}^{(1)})(r_{k}^{(2)}-r_{k,B}^{(2)})}=\sqrt{r_{k}^{(1)}r_{k}^{(2)}}-\sqrt{r_{k,B}^{(1)}r_{k,B}^{(2)}}.
\end{equation}
By combining Eq. (63) and Ineq. (62) above, we conclude that
\begin{equation}
\sqrt{(1-\sum_{k=1}^{m}r_{k,A}^{(1)})(1-\sum_{k=1}^{m}r_{k,A}^{(2)})}-|\alpha|+\sum_{k=1}^{m}\sqrt{r_{k,A}^{(1)}r_{k,A}^{(2)}}|\alpha|^{k+1}\geq
0.
\end{equation}
Due to the above conditions, Ineq. (64) and Ineq. (37) are
equivalent, and, therefore, the proof has been completed. $\Box$

{\it Remark 1.}  Theorem 1 shows that the two-step cloning
protocol in terms of classical one-way or two-way communication
can achieve the optimal efficiency by the NCMSI. This theorem
generalizes Theorem 2 of [38]. Indeed, for $i=1,2$, given integer
$m>0$, if we take $r_{k}^{(i)}=0$ for any $k\not=m$, then from the
above proof we can also take $r_{k,B}^{(i)}=0$, and
$r_{k,A}^{(i)}=0$ for any $k\not=m$. In this case, Theorem 1
reduces to Theorem 2 of [38] as stated in the beginning of this
section. As well, due to $\lim_{m\rightarrow
\infty}\langle\psi_{i}|\psi_{j}\rangle^{m}=0$ for any $i\not=j$,
when $m\rightarrow \infty$ the unitary transformation
\begin{equation}
U(|\psi_{i}\rangle|\phi_{i}\rangle|P_{0}\rangle)=\sqrt{r_{m}^{(i)}}|\psi_{i}\rangle^{\otimes(m+1)}|0\rangle^{\otimes(m-k)}|P_{m}^{(i)}\rangle
+\sum_{l=m+1}^{N}\sqrt{f_{l}^{(i)}}|\Psi_{l}\rangle_{AB}|P_{l}\rangle
\end{equation}
carries out the unambiguous discrimination of the set
$\{|\psi_{1}\rangle|\phi_{1}\rangle,|\psi_{2}\rangle|\phi_{2}\rangle\}$.
Indeed, firstly, if $|\phi_{1}\rangle$ and $|\phi_{2}\rangle$ are
orthogonal, then in Ineq. (36) we take $r_{m}^{(1)}=r_{m}^{(2)}=1$
and $p_{m}=0$, which is in accord with the result that
$\{|\psi_{1}\rangle|\phi_{1}\rangle,|\psi_{2}\rangle|\phi_{2}\rangle\}$
can be exactly discriminated thanks to the orthogonality. If
$|\phi_{1}\rangle$ and $|\phi_{2}\rangle$ are nonorthogonal, then
$|\beta|>0$, and we can take $m$ big enough such that
$|\beta|>|\alpha|^{m}$. Therefore, by using Ineq. (36) we have
that
\begin{equation}
\frac{r_{m}^{(1)}+r_{m}^{(2)}}{2}\leq
\frac{1-|\alpha\beta|}{1-|\alpha|^{m}|p_{m}|}.
\end{equation}
By taking $p_{m}=0$ we obtain that
\begin{equation}
\frac{r_{m}^{(1)}+r_{m}^{(2)}}{2}\leq 1-|\alpha\beta|.
\end{equation}
This has been dealt with by Chen and Yang [39] for achieving the
optimal unambiguous discrimination of any two nonorthogonal pure
product multipartite states with any {\it a priori} probabilities
via local operation and classical communication.

Next we may ask whether or not the two-step protocol is strictly
stronger than the NCMSI. By the following Theorem 2 we show that
the optimal efficiency obtained by the above two-step cloning
protocol can also be achieved by some NCMSI. Therefore, they
indeed have the same optimal efficiency.

{\it Theorem 2.}--For any unitary operators $U_{A}$ and $U_{B}$
satisfying Eqs. (29,31), there is a unitary operator $U$
satisfying Eq. (27), such that
\begin{equation}
r_{k}^{(i)}=
r_{k,B}^{(i)}+\left(1-\sum_{l=1}^{m}r_{l,B}^{(i)}\right)r_{k,B}^{(i)},
\end{equation}
for $k=1,2,\ldots,m$ and $i=1,2$.

{\it Proof:} Leave $\alpha$ and $\beta$ as they are. If
$|\beta|\leq\sum_{k=1}^{m}\sqrt{r_{k}^{(1)}r_{k}^{(2)}}|\alpha|^{k}$,
where $r_{k}^{(i)}=
r_{k,B}^{(i)}+(1-\sum_{l=1}^{m}r_{l,B}^{(i)})r_{k,B}^{(i)}$, then
Ineq. (36) is always satisfied by taking appropriate $p_{k}$,
i.e., the states $|P_{k}^{(i)}\rangle$ of the probe device for
$k=1,2,\ldots,m$ and $i=1,2$.  Hence, we assume that
$|\beta|>\sum_{k=1}^{m}\sqrt{r_{k}^{(1)}r_{k}^{(2)}}|\alpha|^{k}$,
in the following. First we note that
\begin{eqnarray}
\nonumber&&\sqrt{(1-\sum_{k=1}^{m}r_{k,A}^{(1)})(1-\sum_{k=1}^{m}r_{k,A}^{(2)})}\sqrt{(1-\sum_{k=1}^{m}r_{k,B}^{(1)})(1-\sum_{k=1}^{m}r_{k,B}^{(2)})}\\
&=&\sqrt{(1-\sum_{k=1}^{m}r_{k}^{(1)})(1-\sum_{k=1}^{m}r_{k}^{(2)})}.
\end{eqnarray}
Since
$|\beta|>\sum_{k=1}^{m}\sqrt{r_{k}^{(1)}r_{k}^{(2)}}|\alpha|^{k}$,
Ineqs. (40,41) hold, and by these two inequalities, we have
\begin{eqnarray}
\nonumber&&\sqrt{(1-\sum_{k=1}^{m}r_{k,A}^{(1)})(1-\sum_{k=1}^{m}r_{k,A}^{(2)})}\sqrt{(1-\sum_{k=1}^{m}r_{k,B}^{(1)})(1-\sum_{k=1}^{m}r_{k,B}^{(2)})}\\
\nonumber&\geq&(|\alpha|-\sum_{k=1}^{m}\sqrt{r_{k,A}^{(1)}r_{k,A}^{(2)}}|\alpha|^{k+1})(|\beta|-\sum_{k=1}^{m}\sqrt{r_{k,B}^{(1)}r_{k,B}^{(2)}}|\alpha|^{k})\\
\nonumber
&=&|\alpha\beta|-\sum_{k=1}^{m}\sqrt{r_{k,B}^{(1)}r_{k,B}^{(2)}}|\alpha|^{k+1}-|\beta|\sum_{k=1}^{m}\sqrt{r_{k,A}^{(1)}r_{k,A}^{(2)}}|\alpha|^{k+1}\\
&&+\left(\sum_{k=1}^{m}\sqrt{r_{k,A}^{(1)}r_{k,A}^{(2)}}|\alpha|^{k+1}\right)\left(\sum_{k=1}^{m}\sqrt{r_{k,B}^{(1)}r_{k,B}^{(2)}}|\alpha|^{k}\right).
\end{eqnarray}
Therefore, to show Ineq. (39), it suffices to verify that
\begin{eqnarray}
\nonumber &&(|\alpha|-\sum_{k=1}^{m}\sqrt{r_{k,A}^{(1)}r_{k,A}^{(2)}}|\alpha|^{k+1})(|\beta|-\sum_{k=1}^{m}\sqrt{r_{k,B}^{(1)}r_{k,B}^{(2)}}|\alpha|^{k})\\
&\geq&|\alpha\beta|-\sum_{k=1}^{m}\sqrt{r_{k}^{(1)}r_{k}^{(2)}}|\alpha|^{k+1}.
\end{eqnarray}
In terms of Eq. (70), Ineq. (71) is equivalent to
\begin{eqnarray}
\nonumber \sum_{k=1}^{m}\sqrt{r_{k}^{(1)}r_{k}^{(2)}}|\alpha|^{k}
&\geq&\sum_{k=1}^{m}\sqrt{r_{k,B}^{(1)}r_{k,B}^{(2)}}|\alpha|^{k}+|\beta|\sum_{k=1}^{m}\sqrt{r_{k,A}^{(1)}r_{k,A}^{(2)}}|\alpha|^{k}\\
&&-\sum_{k=1}^{m}\sqrt{r_{k,A}^{(1)}r_{k,A}^{(2)}}|\alpha|^{k}\sum_{k=1}^{m}\sqrt{r_{k,B}^{(1)}r_{k,B}^{(2)}}|\alpha|^{k}.
\end{eqnarray}
By using Ineq. (41), it is enough to show that
\begin{eqnarray}
\nonumber \sum_{k=1}^{m}\sqrt{r_{k}^{(1)}r_{k}^{(2)}}|\alpha|^{k}
&\geq&\sum_{k=1}^{m}\sqrt{r_{k,B}^{(1)}r_{k,B}^{(2)}}|\alpha|^{k}\\
&&+\sqrt{(1-\sum_{k=1}^{m}r_{k,B}^{(1)})(1-\sum_{k=1}^{m}r_{k,B}^{(2)})}
\sum_{k=1}^{m}\sqrt{r_{k,A}^{(1)}r_{k,A}^{(2)}}|\alpha|^{k}.
\end{eqnarray}
We can easily check that for any $k=1,2,\ldots,m$,
\begin{equation}
\sqrt{r_{k}^{(1)}r_{k}^{(2)}}\geq
\sqrt{(1-\sum_{l=1}^{m}r_{l,B}^{(1)})(1-\sum_{l=1}^{m}r_{l,B}^{(2)})}\sqrt{r_{k,A}^{(1)}r_{k,A}^{(2)}}+\sqrt{r_{k,B}^{(1)}r_{k,B}^{(2)}},
\end{equation}
which follows from the inequality
\begin{eqnarray}
&&\nonumber r_{k,B}^{(1)}(1-\sum_{k=1}^{m}r_{k,B}^{(2)})r_{k,A}^{(2)}+r_{k,B}^{(2)}(1-\sum_{k=1}^{m}r_{k,B}^{(1)})r_{k,A}^{(1)}\\
&\geq&2\sqrt{(1-\sum_{k=1}^{m}r_{k,B}^{(1)})(1-\sum_{k=1}^{m}r_{k,B}^{(2)})r_{k,B}^{(1)}r_{k,A}^{(2)}r_{k,B}^{(2)}r_{k,A}^{(1)}}.
\end{eqnarray}
Therefore, we complete the proof. $\Box$

{\it Remark 2.} Since cloning only {\it one} multiple copies is a
special case of cloning superposition of multiple clones, Theorem
2 above shows that in Theorem 2 of [38], probabilistic cloning
with supplementary information and the two-step cloning protocol
is equivalent. Therefore this completes Theorem 2 of [38].

{\it Remark 3.} If $|\psi_{1}\rangle, |\psi_{2}\rangle$ are
linearly independent, and $|\phi_{1}\rangle, |\phi_{2}\rangle$ are
linearly dependent, then by virtue of Lemma 1 in [38], the success
probability of Bob running the cloning device described by unitary
operator $U_{B}$ is zero. Therefore, in this case, the NCMSI has
the same cloning efficiency as the NCM. However, if the
supplementary information $|\phi_{1}\rangle, |\phi_{2}\rangle$ are
linearly independent, then the success probabilities in the
cloning machine described by $U_{B}$ are likely bigger than zero,
and, thus, from Theorem 2 it follows that the success probability
of the NCMSI for cloning is bigger than the NCM [13].

\section*{4. Concluding remarks}

\hskip 6mm We have dealt with the novel cloning machine with the
help of supplementary information (NCMSI) for producing quantum
superposition of multiple copies. When two holders, say Alice and
Bob, possess respectively the original and the supplementary
information, we have derived that the optimal efficiencies of
cloning achieved via quantum communication and via classical
one-way or two-way communication between the two parties in these
devices are indeed equivalent. Therefore, the NCMSI for producing
quantum superposition of multiple copies may have bigger success
probability than the NCM [13]. However, by classical communication
we do not know how to obtain the all copies together in a quantum
computer, so, in practice we may use the scenario of quantum
communication, i.e., the NCMSI.

As stated in Section 1, probabilistic
cloning may get precise copies with certain probability, so,
improving the success ratio is of importance. We hope that our
results would provide some useful ideas in preserving important
quantum information, parallel storage of quantum information in a
quantum computer, and quantum cryptography.

When cloning $n$ states with $n\geq 3$, Azuma {\it et al.} [38]
demonstrated that the optimal efficiency of copying achieved via
quantum communication between the original and the supplementary
parties sometimes cannot be accomplished by using only classical
channel. Then an interesting problem is what is the sufficient and
necessary condition for retaining the equivalence as we proved in
this paper. A possible method is to combine matrix theory [40] and
the present paper. Moreover, if the supplementary information is
given as a mixed state or we have multiple supplementary
information, then the probabilistic or novel cloning devices are
still worth considering. We would like to explore these questions
in future.

\section*{Acknowledgements}
I am very grateful to the referees for their invaluable comments
and suggestions that help to improve the presentation of this
paper. This work is supported by the National Natural Science
Foundation (No. 90303024, 60573006), the Higher School Doctoral
Subject Foundation of Ministry of Education (No. 20050558015), and
the Natural Science Foundation of Guangdong Province (No. 020146,
031541) of China.

\end{document}